\documentclass[useAMS,usenatbib]{mn2e}
\usepackage{graphicx}
\usepackage{txfonts}
\title[Broad band X-ray spectrum of the newly discovered BLRG IGR J21247+5058]
  {Broad band X-ray spectrum of the newly discovered Broad Line Radio Galaxy IGR J21247+5058}
\author[M. Molina et al.]
{M.~Molina,$^1$ M.~Giroletti,$^2$ A.~Malizia,$^3$ R.~Landi,$^3$ L.~Bassani$^3$, A.J.~Bird,$^1$ A.J.~Dean,$^1$  \\
\newauthor
A.~De Rosa,$^4$ M.~Fiocchi,$^4$ F.~Panessa$^5$\\
$^1$School of Physics and Astronomy, University of Southampton,
        SO17 1BJ, Southampton, U.K., \\
$^2$IRA/INAF, via Gobetti 101, I-40129 Bologna, Italy,\\
$^3$IASF/INAF, via Gobetti 101, I-40129 Bologna, Italy\\
$^4$IASF/INAF, via Fosso del Cavaliere 100, I-00133 Rome, Italy\\
$^5$Instituto de Fõsica de Cantabria (CSIC-UC), Avda. de los Castros, 39005 Santander, Spain }

\begin{document}

\date{}

\pagerange{\pageref{firstpage}--\pageref{lastpage}} \pubyear{2007}

\maketitle

\label{firstpage}
        
\begin{abstract}

In this paper we present radio and high energy observations of the \emph{INTEGRAL} source
IGR J21247+5058, a broad line emitting galaxy obscured by the Galactic plane. Archival VLA radio data
indicate that IGR J21247+5058 can be classified as an FRII Broad Line Radio Galaxy. 
The spectrum between 610 MHz and 15 GHz is typical of synchrotron self-absorbed radiation 
with a peak at 8 GHz and a low energy turnover; the core fraction is 0.1 suggestive of  a moderate
Doppler boosting of the base of the jet. The high energy broad-band spectrum was obtained by 
combining \emph{XMM-Newton} and \emph{Swift/XRT} observation with \emph{INTEGRAL/IBIS} data.
The 0.4-100 keV spectrum is well described by a power law,
with slope $\Gamma$=1.5, characterised by complex absorption due to two layers of material  partially 
covering the source and a high energy cut-off around 70-80 keV.  Features such as a narrow iron line and a 
Compton reflection component, if present, are weak, suggesting that reprocessing of the power law 
photons in the accretion disk plays a negligible role in the source. 

\end{abstract}


\section{Introduction}

IGR J21247+5058 was initially reported in the first \emph{INTEGRAL} survey
catalogue \citep{b5} and listed in subsequent survey papers: in the
most recent work the source is located at RA (J2000)=321.172 and Dec
(J2000)=+50.972 with an associated 90\% error circle of 1$^{\prime}$
\citep{b7}.  Soon after its discovery, it was associated by
\citet{b23} with the radio source 4C 50.55, also known as GPSR
93.319+0.394, KR2, NRAO 659 or BG 2122+50: this object has the typical
morphology of a radio galaxy showing a bright core
and two lobes. The estimated position of the core from the NVSS map 
is RA (J2000)=21$^{h}$24$^{m}$39.25$^{s}$ and Dec (J2000)=+50$^{\circ}$58$^{\prime}$23.80$^{\prime\prime}$  
(1$^{\prime\prime}$ uncertainty).

Confirmation of the AGN nature of IGR J21247+5058 came via optical
observations obtained at the Loiano telescope \citep{b21}, despite the
fact that the optical spectrum of the source looks very peculiar. It
has in fact a broad, redshifted H$_{\alpha}$ complex superimposed onto
a {\textquotedblleft{normal}\textquotedblright} F/G-type Galactic star
continuum. While most of the observed features (Na, Ca and Mg) are
consistent with redshift z=0 and thus with a Galactic stellar origin,
the H$_{\alpha}$ complex leads to a redshift z=0.02. This feature is very similar 
to the one observed in another bright radio galaxy, namely 3C390.3 
\citep{b72}. The H$_{\alpha}$ complex,
together with the spatially coincident extended radio emission and
the detection of strong hard X-ray radiation, strongly indicates the
unfortunate situation of a chance alignment between a relatively
nearby star and a background radio galaxy. Indeed the
\emph{INTEGRAL/IBIS} spectrum of this source is compatible with the
canonical AGN spectrum \citep{b22}. Because the only optical line
observed is broad, the source was tentatively classified as a Seyfert
1 or alternatively as a broad line radio galaxy (BLRG). Unfortunately,
given the confusion with the nearby star, it is impossible to gain more information on the 
source optical characteristics with the currently available data (see \citealt{b21}); 
in particular no reliable measurement of the B magnitude can be used to estimate the 
source radio loudness using the relation RL=log[F(5GHZ)/F(B)].
At the observed redshift, the source luminosity in the 20-100 keV
band is 8.5$\times$10$^{43}$ erg s$^{-1}$, making IGR J21247+5058
one of the brightest AGN in the local Universe\footnote{Assuming H$_0$=70
km s$^{-1}$Mpc$^{-1}$ and a flat Universe.}.

Here, we present a detailed radio analysis of the source
based on archival VLA data. We also discuss archival \emph{XMM} and 
\emph{Swift}/\emph{XRT} data in combination with a new
\emph{INTEGRAL/IBIS} spectrum which capitalises on the larger exposure
now available on this source. 

\section{Radio Observations}

At the very low energy part of the electromagnetic spectrum, radio observations
provide valuable information on the nature of IGR J21247+5058. 
Radio images at various resolutions have been presented in several works
(e.g. \citealt{b20}, \citealt{b42}). Here we show in Figure \ref{1.4GHz} a 1.4 GHz image 
of the field of IGR J21247+5058 obtained with data
from the VLA archive\footnote{The National Radio Astronomy Observatory is a facility of the
National Science Foundation operated under cooperative agreeement by Associated Universities, Inc.}. 
The circle indicates the location and positional error of
the \emph{INTEGRAL} detection, which clearly points to the nucleus of IGR J21247+5058 
as the source of the gamma-ray emission.
The radio source has the typical edge-brightened
morphology of an FRII radio galaxy, with a central compact core and two large
lobes. The size of IGR J21247+5058 is $\sim9.5^{\prime}$ and the total flux density at 1.4
GHz is 2.5 Jy. At the redshift proposed for this source, 
these data correspond to a total extent LS=230 kpc and a
monochromatic radio power $P_\mathrm{1.4}=10^{24.4}$W Hz$^{-1}$. This makes
IGR J21247+5058 a typical radio galaxy in size, with a radio power intermediate
between FRI and FRIIs.

In Figure \ref{av_spec} we show the spectrum of the core of IGR J21247+5058 between 610 MHz and 15
GHz. The data between 1.4 and 15 GHz were obtained from data in the VLA
archive, while the 610 MHz point is taken from the GMRT data \citep{b42}.
The spectrum is typical of synchrotron self-absorbed radiation, with a
peak at about 8 GHz and a low frequency turnover. 

The core fraction at 1.4 GHz is about $S_c/S_t=0.1$, which is suggestive of a
moderate Doppler boosting of the base of the jet. 
In fact, the core is brighter
than what would be expected on the basis of the correlation between core and total radio
power \citep{b43}. From the total flux density at low frequency
($S_{0.4}=5.4$ Jy, \citealt{b20}), the core flux density should be
only $\sim 50$ mJy at 5 GHz, i.e. about a factor 10 less than observed. If we
assume a typical Lorentz factor for the radio jet ($\gamma$=5, see e.g. \citealt{b44}), 
we can use this constraint to estimate a viewing angle
$\theta \sim 35^{\circ}$. This seems to be small enough to allow us to peer into the
BLR and it also explains the broadening of the H$_{\alpha}$ line. 
Put altogether, the radio data seem to indicate
that the counterpart of IGR J21247+5058 is an FRII broad line radio galaxy.

It is however difficult to guess which is the approaching side of the
source. The NW lobe is brighter, but a knot of enhanced brightness is visible
in the SE jet at 90$^{\prime\prime}$ from the core. If this brightness asymmetry is due to
Doppler boosting, then the approaching side would be the SE one. A look at the
parsec scale structure would be desirable to better study the properties of the
inner jet and define this issue.

It is also interesting to note that a weak feature is clearly detected in
several VLA data sets at $\sim 2^{\prime}$ south of the core (RA = 21$^h$24$^m$39.97$^s$, Dec =
+50$^{\circ}$56$^{\prime}$05.4$^{\prime\prime}$). It has a flux of $\sim 4$ mJy at 1.4 GHz and $\sim2$ mJy at 4.8
GHz. This source is probably unrelated to the \emph{INTEGRAL} source, as it falls
outside its 90\% error circle and it is not detected in X-rays (see next section).

\begin{small}
\begin{figure}
\centering
\includegraphics[width=0.8\linewidth]{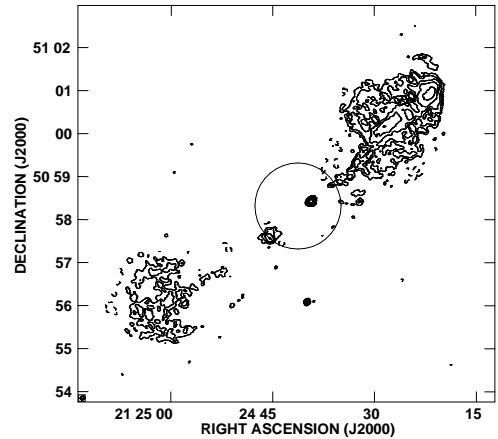}
\caption{A 1.4 GHz image of the field of IGR J21247+5058 from VLA
data (in B configuration). Contours are traced at ($-1$, 1, 2, 4, ...)$\times$0.5 mJy beam$^{-1}$
and the peak is 244 mJy beam$^{-1}$. The restoring beam is 5.5$\times$4.4
arcsec. The circle shows the ISGRI position and error box of IGR J21247+5058.}
\label{1.4GHz}
\end{figure}
\end{small}

\begin{small}
\begin{figure}
\centering
\includegraphics[width=0.8\linewidth]{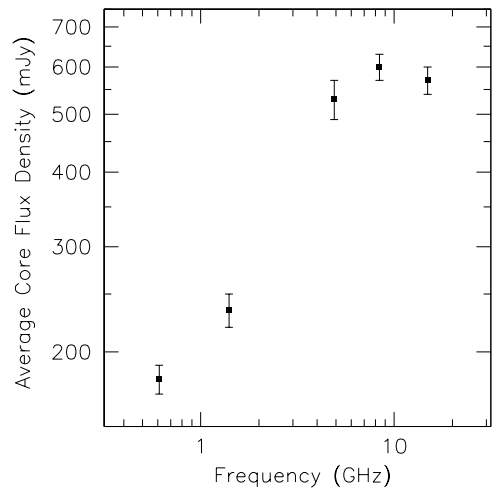}
\caption{Average spectrum of the core of IGR J21247+5058 with VLA data (between 1.4 and
15 GHz) and the GMRT (0.6 GHz). The error bars show the rms of the various
measurements considered, as well as the absolute calibration uncertainty.}
\label{av_spec}
\end{figure}
\end{small}

\section{X-ray observations and data analysis}

IGR J21247+5058 was observed by \emph{XMM-Newton} on 2005 November 6
during orbit 1083, in the \emph{XMM-Newton} Guest Observer Programme. Two other 
observations are present in the archive, but have not been used due to their poor statistical 
quality. During this orbit, the EPIC PN \citep{b24} exposure was $\sim$25
ks, while the EPIC MOS1 and MOS2 \citep{b25} exposures were
$\sim$27 ks. The EPIC PN camera was operated in Large Window Mode with
a thick filter applied, while the two MOS cameras were both operated
in Small Window Mode.
MOS and PN data were reprocessed using the \emph{XMM-Newton} Standard
Analysis Software (SAS) version 7.0. Image analysis indicates
that a bright source, localised at RA=21$^{h}$24$^{m}$39.36$^{s}$ and 
Dec (J2000)=+50$^{\circ}$58$^{\prime}$23.86$^{\prime\prime}$,
compatible with the position of the radio core,
is detected with high significance, while no emission is seen at the 
location of the radio lobes nor in correspondence with 
the radio source located south of the core of IGR J21247+5058. 
As for the Galactic star which is aligned with the radio galaxy by chance, it could emit
X-rays by coronal activity as seen in many stars of similar late spectroscopic type.
However, any contamination is likely to be negligible as the star is likely at 2.5 Kpc \citep{b21}, 
i.e. too far away to provide significant flux if the emission is from a stellar corona.
MOS (PN) source counts were extracted from circular regions of 50$^{\prime\prime}$ radius 
centered on the source, while background spectra were
extracted from source free regions of 20$^{\prime\prime}$ radius. 
Pile up is negligible for each of the instruments.
In this analysis, only patterns corresponding to single, double, triple and quadruple
events for the two MOS cameras were selected (PATTERN$\leq$12), while
for the PN only single and double events (PATTERN$\leq$4) were taken
into account; the standard selection filter FLAG=0 was applied.  
Exposures have been filtered for periods of high background activity and the
resulting values are listed in Table 1, together with other relevant information related to 
this \emph{XMM} measurement.

\begin{table*}
\begin{center}
\centerline{{\bf Table 1}}
\vspace{0.2cm}
\begin{tabular}{lccccc}
\multicolumn{5}{c}{{\bf  Observations Log}}\\
\hline
\hline
                                     & {\bf Obs date}          &  {\bf Exposures (ksec)} & {\bf Filter}   &    {\bf Source counts (ct/s)}\\
\hline
\emph{XMM-MOS1} &  06-11-2005       &            24.2            &   Thick  &    1.918$\pm$0.009\\
\emph{XMM-MOS2}  &  06-11-2005      &            23.6            &   Thick   &   1.907$\pm$0.009\\
\emph{XMM-PN}       &  06-11-2005       &            22.0            &   Thick   &   5.582$\pm$0.02\\
\emph{XRT1}             &  17-10-2006       &             7.2             &    -          &   0.4953$\pm$0.01\\
\emph{XRT2}             &  25-01-2007       &            3.7              &    -           &   0.3222$\pm$0.01\\
\emph{INTEGRAL}   & Nov. 2002 to Apr. 2006 &         768                &    -           &   1.423$\pm$0.04\\
\hline
\end{tabular}
\end{center}
\end{table*}

The ancillary response
matrices (ARFs) and the detector response matrices (RMFs) were
generated using the \emph{XMM}-SAS tasks \emph{arfgen} and
\emph{rmfgen}; spectral channels were rebinned in order to achieve a
minimum of 20 counts per each bin. Here and in the following, spectral
analysis was performed with XSPEC v.11.2.3 \citep{b1} and errors are
quoted at 90\% confidence level for one parameter of interest
($\Delta\chi^{2}$=2.71). Since the source is located behind the Galactic plane,
Galactic absorption is high, being 1.11$\times$10$^{22}$atoms cm$^{-2}$ \citep{b61}; 
it has therefore been included in each fit so that the quoted absorbing column densities
are always in excess to this Galactic value.
Cross calibration constants PN/MOS1 and MOS2/MOS1 were left free to vary 
and always found to be close to unity as expected.
Data from the three EPIC cameras were initially fitted in the 0.4-10
keV range using a simple power law absorbed only by the Galactic column density
(see Figure \ref{xmm_pl}). This model does not yield an acceptable fit 
($\chi^2$=4974.7 for 2624) as evident in Figure \ref{xmm_pl} and the resulting power law slope is very flat 
($\Gamma$=0.94$^{+0.01}_{-0.01}$). Since the data to model ratio are indicative of intrinsic absorption, the
data were re-fitted adding this component (\texttt{wa$_g$*wa*po} in XSPEC); 
this new model provides a significant improvement in the fit ($\chi^2$=2526.2 for 2623 d.o.f.),
a column density N$_H$=0.62$^{+0.02}_{-0.02}$$\times10^{22}$cm$^{-2}$ but a still rather flat
spectrum ($\Gamma$=1.28$^{+0.01}_{-0.01}$). An even better fit ($\chi^2$=2446.1 for 2622 dof)
is obtained using an ionized absorber instead of a cold one (\texttt{wa$_g$*absori*po} in XSPEC): 
the column density in this case is 1.12$^{+0.10}_{-0.09}$$\times10^{22}$cm$^{-2}$ and the  
ionization state  $\xi$ (= L/nR$^2$, \citealt{b71}) is 18.1$^{+6.8}_{-5.9}$, i.e. the absorber is at most 
mildly ionized. The slope of the power law hardens but only marginally ($\Gamma$=1.33$^{+0.02}_{-0.02}$). 
We then substituted the absorption (cold or ionized) with a partial covering component
(\texttt{*wa$_g$*pcfabs*po}, see Table 2). This model
provides a $\Delta\chi^2$=103.8 for 1 d.o.f. compared to the cold absorption model 
and a better $\chi^2$ for the same d.o.f. compared to the ionized absorption model; 
the resulting spectrum has a similar photon index and column density as obtained for the ionized absorber model, 
but the absorption is now colder and covers around 80\% of the source. At this stage we also
checked the data for the presence of a cold iron line, adding to the partially covering model
a narrow Gaussian component, having fixed the width to 10 eV. This model provides a fit improvement
($\chi^2$=2405.9 for 2620 d.o.f. or $\Delta\chi^2$=16.5 for 2 d.o.f.), a line energy at 
6.39$^{+0.06}_{-0.08}$ keV and an equivalent width (EW) of 20.6$^{+6.1}_{-9.6}$ eV (see Figure \ref{line_cont}
for the confidence countours of the line energy versus line normalisation). The improvement is significative at the 
99.9\% confidence level; however the EW is rather small, i.e. too close to the capability limits of 
moderate resolution CCD instruments like \emph{XMM}, calling for some caution in considering the line as a real feature or just local noise. 
Without going into more details, the important point to stress here is that in IGR J21247+5058 a cold iron  
line is either very weak or not present.
There is also a hint for a line at around 1.7 keV, but again its EW is very small and its energy suspiciously 
close to background features present in the PN and MOS camera to consider it as a real line.
 
\begin{small}
\begin{figure}
\centering
\includegraphics[width=0.7\linewidth, angle=-90]{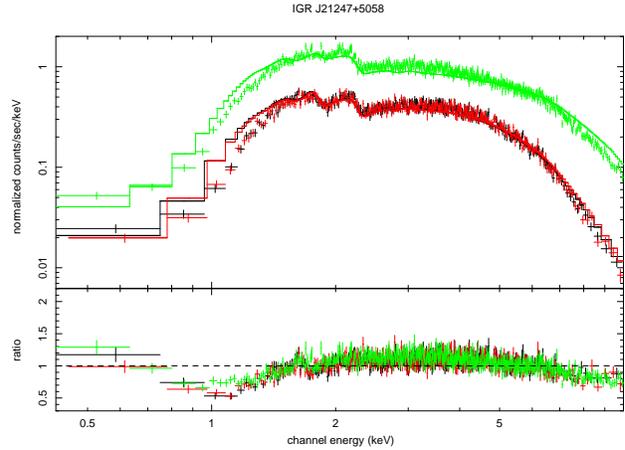}
\caption{The \emph{XMM} data fitted using a simple power law absorbed by Galactic column density.}
\label{xmm_pl}
\end{figure}
\end{small}

\begin{small}
\begin{figure}
\centering
\includegraphics[width=0.7\linewidth, angle=-90]{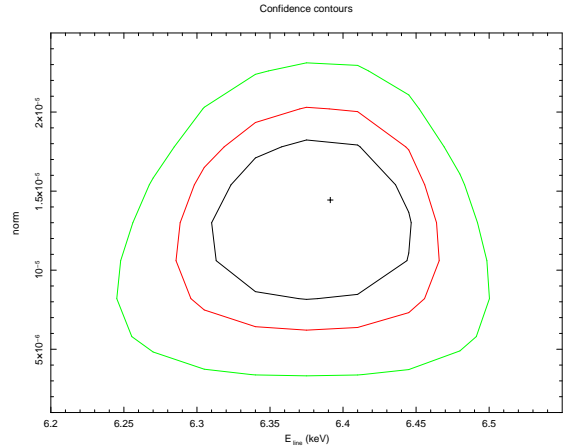}
\caption{The confidence contours of the line energy versus line normalisation.}
\label{line_cont}
\end{figure}
\end{small}

Despite the fact that a fit with just one absorber (cold, mildly ionized or partially covering the souce)
is quite acceptable, the resulting spectrum is still rather flat. Hence we tried two different scenarios 
to steepen the \emph{XMM} spectral data: reflection from neutral material (with the Compton reflection component described
by the parameter \emph{R}; \texttt{pexrav} model in XSPEC) or an extra layer of absorbing material (again
cold, ionized or partially covering the source). In the first case the inclination angle was fixed at 30$^{\circ}$, implying a nearly face-on 
geometry. The reflection component resulting from the fit was small R=0.59$^{+0.45}_{-0.42}$ and the photon index still rather flat 
($\Gamma$=1.43$^{+0.06}_{-0.06}$); the low value of the reflection parameter, although not well constrained,
is consistent with the weak iron line observed in the 
spectrum of IGR J21247+5058, suggesting that the reprocessing of the power law photons in the accretion disk plays a negligible role
in the source. In the second case,
the best fit is obtained with another layer of cold material partially covering the source
(\texttt{wa*pcfabs*pcfabs*po}, see Table 2). This model 
provides a power law slope of $\sim$1.6, close to the 1.7 value  measured in other BLRG \citep{b33, b39}, 
although still flatter than the 1.9 value typically observed in radio quite AGN \citep{b29}. 
The absorbing column densities  
are around  9$\times10^{22}$ cm$^{-2}$ and 10$^{22}$cm$^{-2}$,
covering 27\% and 83\% of the central source. Given the fact that the highest steepening of the spectrum is obtained 
with a double absorber model and that reflection is either not required  or small (also from the weakness of the iron line),
we chose the first model as a better description of the \emph{XMM} data.
As some excess counts (mostly PN) are still visible at low energies,  
we have further added to this best fit model a soft thermal component (in the form of a black body or a \texttt{mekal} model)
or a scattered component in the form of a power law model; 
in the latter case the two power laws have the same photon index but different normalisations. In all cases,
the excess emission remains and the fits return parameters which are unusual for AGN (a too high temperature 
and a scattered component with a too high normalisation with respect to the primary continuum). 
Given the quality of the data at low energies and the small strength of this excess emission, 
it is beyond the scope of the present paper to further enquire about this component
and its analysis is postponed to when more detailed X-ray observations of IGR J21247+5058 are available.

In order to check if the source could be classified as a
radio-loud object, we exploited the X-ray flux measurement and 
used the R$_X$=L$_r$(5GHz)/L$_X$(2-10 keV) relation 
(e.g. \citealt{b26} and \citealt{b36});  to do this we used the VLA 5 GHz 
measurement of the core component and then compared it to the 2-10 keV flux.
Since LogR$_X$=-3.3, IGR J21247+5058 could be defined a borderline object, as it is radio loud
or quiet depending on our choice of the dividing line between these two classes: LogR$_X$=-4.5
(as in \citealt{b26}) or -2.8 (as in \citealt{b36}). However the LogR$_X$ value of IGR J21247+5058 is fully 
compatible with those of similar BLRG like 3C 111 (LogR$_X$=-3.7), 3C120 (LogR$_X$=-2.1), 3C3 90.3 (LogR$_X$=-3.1) 
and 3C 382 (LogR$_X$=-4.0) \citep{b39, b67},
implying that also this new source is most likely a radio loud AGN.

IGR J21247+5058 was also observed with the \emph{XRT} (X-ray Telescope,
operating in the 0.2-10 keV range) on board the \emph{Swift} satellite
\citep{b15} for $\sim$7.2 ks on 2006 October 17 and for 3.7 ks on 2007
January 25.  Data reduction was performed using the XRTDAS v1.8.0
standard data pipeline package ({\sc xrtpipeline} v. 0.10.3) in order
to produce screened event files. All data were collected in the
Photon Counting (PC) mode \citep{b17}, adopting the standard grade
filtering (0-12 for PC) according to the \emph{XRT} nomenclature.  
Source data have been extracted using photons in a circular region of
radius 20$^{\prime\prime}$; background data have instead been taken
from various uncontaminated regions near the X-ray source,
using either a circular region of different radius or an annulus
surrounding the source. The log of these two \emph{XRT} observations 
is also reported in Table 1.

Due to the lower quality of these data, we have employed a simple model, i.e. a
power law passing through a single aborption layer partially covering the central source.
The first \emph{XRT} observation (hereafter \emph{XRT1}) provides a power law spectrum
with $\Gamma$=1.54$^{+0.19}_{-0.18}$, moderate intrinsic absorption,
a partial covering fraction \emph{f}=0.74 and a 2-10 keV flux of 6.4$\times$10$^{-11}$ erg cm$^{-2}$s$^{-1}$
(see Table 1). To check if the best fit values could be improved, we added another partial 
covering absorber to the model, but the fit does not produce better results 
($\chi^2$=132.2 for 138 d.o.f.). 

The second observation (hereafter \emph{XRT2}) is of even poorer quality due to the lower
exposure; it provides slightly different parameters than the first observation (see Table 2)
and a lower 2-10 keV flux of (3.37$\pm0.1$)$\times$10$^{-11}$ erg cm$^{-2}$ s$^{-1}$. 
Despite the less precise modelling allowed by the \emph{XRT} observations, the comparison 
between the two \emph{XRT} observations and with the \emph{XMM-Newton} measurement suggests a change 
in the source absorption properties related to a flux variation. Note that for consistency,
also the \emph{XMM} data are modelled here with a single layer of absorption.
The change in the absorber properties is evident in Figure \ref{cont_pcfabs} where the contour plots of 
the column density versus covering fraction are displayed for the three available X-ray observations: in particular
while \emph{XRT1} data are fully compatible with the \emph{XMM} one, both differ from the \emph{XRT2} observation parameters.
Given that the model used is not the best fit found with the \emph{XMM} data, as a second step we adopted the double absorber 
model and  compared the 3 sets of data by means of an accurate parameter space expoloration. 
Due to the lower statistical quality of the \emph{XRT} data, we have fixed to the \emph{XMM} value
the photon index and the two absorption parameters each at a time. 
The result of this procedure is that the likely change occuring is in the column density of the absorber covering 80\% of the source 
while the other parameters are consistent within errors with each other. However, given the poor quality fo the \emph{XRT2} 
spectrum, i.e. the one that provides evidence for this change in absorption, some caution is needed and further observations required to confirm  
this observational evidence.
 
The \emph{INTEGRAL} data reported here consist of several pointings of
IGR J21247+5058 performed by the \emph{IBIS/ISGRI} instrument between
revolution 12 and 429, i.e. the period from launch to the end of April
2006, and correspond to a total exposure of 768 ks. \emph{ISGRI}
images for each available pointing were generated in various energy
bands using the ISDC offline scientific analysis software OSA
\citep{b16} version 5.1. Count rates at the position of the source
were extracted from individual images in order to provide light curves
in various energy bands; from these light curves, average fluxes were
then extracted and combined to produce a source spectrum. Analysis
was performed in the 17-100 keV band (the source count rate in this band is reported in Table 1). 
A simple power law provides a good fit to the \emph{IBIS} data ($\chi^2$=8.9 for 10 d.o.f.)
and a photon index $\Gamma$=2.0$\pm$0.1 and a 17-100 keV flux
of 1.15$\times$10$^{-10}$ erg cm$^{-2}$ s$^{-1}$. The \emph{INTEGRAL} spectrum
is clearly steeper than the \emph{XMM/XRT} ones, implying that a cut-off is possibly present
nd located in the \emph{IBIS} energy band. Up to now IGR J21247+5058 
has not been reported by \emph{Swift}/BAT,
and therefore the IBIS measurement is the only available above 10 keV. 

\begin{small}
\begin{figure}
\centering
\includegraphics[width=0.75\linewidth, angle=-90]{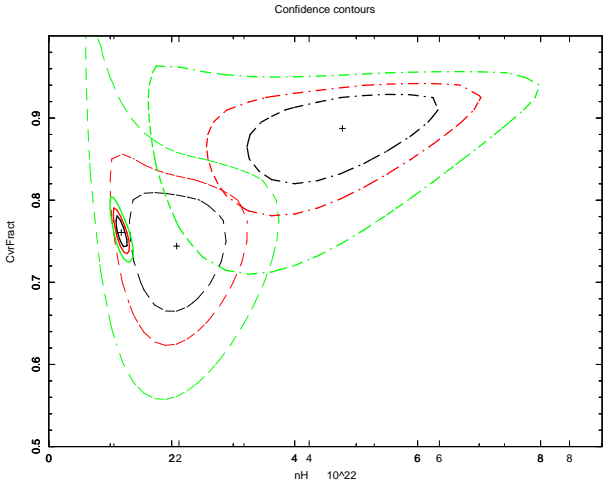}
\caption{Constraints on the absorbing column densities and relative covering fractions for the \emph{XMM} (solid line), 
\emph{XRT1} (dashed line) and \emph{XRT2} (dashed-dotted line) observations assuming a single 
absorption layer (\texttt{wa$_g$*pcfabs*po}).}
\label{cont_pcfabs}
\end{figure}
\end{small}

\subsection{The high energy broad-band spectrum}

X-ray and \emph{INTEGRAL} data were then fitted together  in
order to obtain an average broad-band spectrum of the source.
In the fitting procedure, a multiplicative constant, \emph{C}, has been introduced to take
into account possible cross-calibration mismatches between the X-ray and \emph{INTEGRAL} data; this constant has 
been found to be close to 1 both for \emph{XMM} and \emph{Swift/XRT} using various source typology  \citep{b69, b68, b70}
so that significant deviation from this value can be confidently ascribed to source flux variability.
Initially, we combined the \emph{XMM} and the \emph{IBIS/ISGRI}
data together, employing the model (\texttt{wa*pcfabs*pacfabs*po}) used for the 
\emph{XMM} data alone and considered here as our baseline model (we have ignored the iron line at this stage
but the results do not change significantly if this component is added to the fit).
This model provides a good fit with a photon index of $\sim$1.7, and two absorption layers
($\sim10^{22}$ and  $\sim10^{23}$ cm$^{-2}$ respectively, Table 3),
covering 84\% and 34\% of the central source.
To check for the presence of a high energy cut-off, we substituted 
the simple power law with a cut-off power law (\texttt{wa*pcfabs*pcfabs*cutoffpl}).
The model yields our best fit ($\chi^2$=2354.0, 2628 d.o.f., Figure \ref{xmm_isgri}) and a power law slope
around 1.5; a cut-off is indeed present and well constrained at around 
75 keV (see Table 3 and Figure \ref{cont_cutoff}). In the above models 
the value of \emph{C} is around 0.80, suggesting a good match and no major changes in flux between 
the \emph{XMM} and \emph{INTEGRAL} observations. 
Finally, we exploited the broader energy coverage check again for the presence of 
reflection in the source spectrum by adding this component to the previous model
(\texttt{wa*pcfabs*pcfabs*pexrav}, see Table 3).
The fit shows an improvement ($\chi^2$=2350.9 for 2627 d.o.f., 99.9\%  confidence level)
with respect to the simple double partial covering model (\texttt{wa*pcfabs*pacfabs*po}); however,
when considering the best-fit model (\texttt{wa*pcfabs*pcfabs*cutoffpl}), there  is
only a marginal improvement in the fit, suggesting that the reflection component is
not strongly required by the data. The model gives a power law slope of $\sim$1.5,
the high energy cut-off is around 100 keV
and the double absorption layer values are well in agreement with those 
obtained with no reflection in the model (R=0). This reflection model 
yields a poorly constrained value of \emph{R} around 0.4 and a cross calibration constant $\sim$0.7.
We have also tried to find evidence for a jet component in the high energy data
introducing a second power law component (\texttt{wa*pcfabs*pcfabs*(cutoffpl+po)}) 
with the photon index of the primary continuum fixed at the canonical 1.7 value
for BLRG. Again the fit doesn't show any improvement ($\chi^2$=2379.2, 2627 dof), 
the power law component used to model the jet emission has a slope of 1.50$^{+0.16}_{-0.33}$ and the high 
energy cut-off is consistent with value found in the best fit (83.5$^{+61.6}_{-29.2}$ keV). The values of the
column densities and their covering fractions are in agreement with those found in the best fit. We can thus conlcude that
the extra jet component is not required by the data and that the souce emission is not jet-dominated.

\begin{small}
\begin{figure}
\centering
\includegraphics[width=0.7\linewidth,angle=-90]{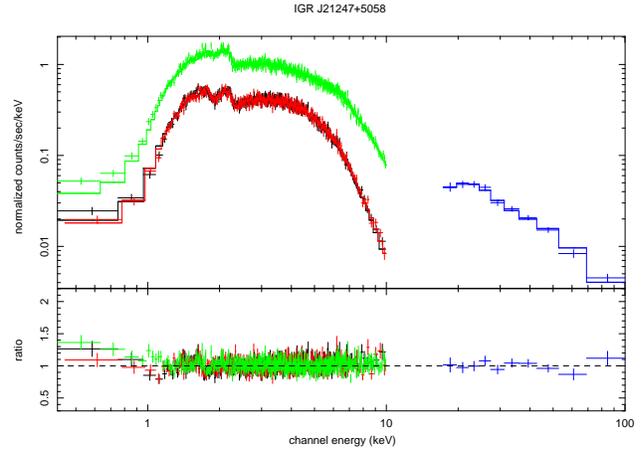}
\caption{\emph{XMM/ISGRI} broad band (0.4-100 keV) spectrum of IGR J21247+5058: the model is a cut-off power-law absorbed
both by Galactic column density and by two layers of absorbing material partially covering the
source.}
\label{xmm_isgri}
\end{figure}
\end{small}

\begin{small}
\begin{figure}
\centering
\includegraphics[width=0.7\linewidth,angle=-90]{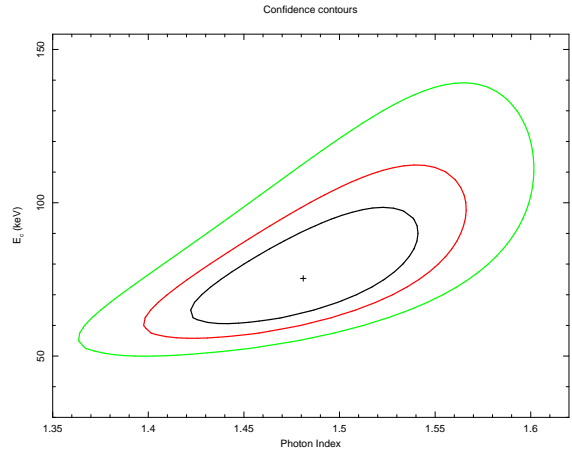}
\caption{Constraints on the primary continuum (power-law photon index) versus the
high energy cut-off of IGR J21247+5058.}
\label{cont_cutoff}
\end{figure}
\end{small}

Next best-fit model (\texttt{wa*pcfabs*pcfabs*cutoffpl}) was used to fit \emph{XRT} and \emph{INTEGRAL} data.
In both \emph{XRT} observations, the fit is good ($\chi^2$=138.8 for 145 d.o.f. and 
$\chi^2$=47.8 for 54 d.o.f.), the power slopes are around 1.6 and 1.8 and the
high energy cut-off values are not well constrained, although they could be placed above 50 and 80 keV respectively.
The partial covering fractions for the \emph{XRT1-2}/\emph {ISGRI} broad-band spectrum
are broadly compatible with those found for the \emph{XMM/ISGRI} spectrum
(see Table 3). It must be pointed out that due to the poor statistical quality of the \emph{XRT} 
datasets, the values of the partial covering fractions are not well constrained and so errors are evaluated 
by freezing the parameters related to each layer while calculating the uncertainties for the other.
The constant between \emph{XRT1}
and \emph{INTEGRAL} spectra is around 0.8, suggesting again  
agreement between the two sets of data. The constant between the 
\emph{XRT2} observation and the \emph{INTEGRAL} data is instead
higher (1.6), implying that the source could have 
undergone some minor changes in flux or have changed its absorption properties as discussed in the previous 
section.

\begin{table*}
\begin{center}
\centerline{{\bf Table 2}}
\vspace{0.2cm}
\begin{tabular}{lccccccc}
\multicolumn{8}{c}{{\bf  Spectral Fits to \emph{XMM-Newton} and \emph{Swift}/\emph{XRT}}}\\
\hline
                                    &{\bf $\Gamma$}              &  {\bf N$_H^1$}                 &       {\bf \emph{f$_1$}}  &  {\bf N$_H^2$}                 & {\bf \emph{f$_2$}}& {\bf F$_{2-10 keV}$}& {\bf $\chi^2$ (dof)}\\
                                   &                                      & {\bf (10$^{22}$cm$^{-2}$)}&                                      & {\bf (10$^{22}$cm$^{-2}$})&                             &{\bf (10$^{-11}$ erg cm$^{-2}$s$^{-1}$)}&                            \\
\hline
{\bf\emph{XMM$^a$}}& 1.35$^{+0.02}_{-0.02}$&  1.11$^{+0.10}_{-0.10}$   & 0.76$^{+0.02}_{-0.02}$ &                      -                    & -                           &  5.1&2422.4 (2622)    \\
{\bf\emph{XMM$^b$}}& 1.57$^{+0.05}_{-0.05}$& 0.99$^{+0.12}_{-0.12}$   & 0.83$^{+0.03}_{-0.02}$ & 9.14$^{+1.93}_{-1.75}$    & 0.27$^{+0.04}_{-0.04}$ &  5.1 & 2345.8 (2620)         \\
{\bf\emph{XRT1$^a$}}& 1.54$^{+0.19}_{-0.18}$ & 2.08$^{+0.88}_{-0.84}$   & 0.74$^{+0.07}_{-0.09}$  & -                                          & -                         &   6.4&     132.5 (140)      \\
{\bf\emph{XRT2$^a$}}& 1.94$^{+0.39}_{-0.37}$ & 4.76$^{+1.73}_{-1.67}$   & 0.89$^{+0.04}_{-0.07}$  &    -                                       & -                         & 3.3&      41.4 (48)    \\
\hline
\end{tabular}
\end{center}
\scriptsize
Best fit parameters related to model (a) (\texttt{wa*pcfabs*po}) and model (b) (\texttt{wa*pcfabs*pcfabs*po});
Parameters are as following: {\bf $\Gamma$}=Photon index; {\bf N$_H^1$}= Column density  of first absorber; \\{\bf \emph{f$_1$}}= Covering fraction of first absorber;  {\bf N$_H^2$}= Column density of second absorber;             {\bf \emph{f$_2$}}= Covering fraction of second absorber; {\bf F$_{2-10 keV}$}= 2-10 keV Flux and {\bf $\chi^2$ (dof)}= Chi squares (degrees of freedom).\\ 
\end{table*}
                           
\begin{table*}
\begin{center}
\centerline{{\bf Table 3}}
\vspace{0.2cm}
\begin{tabular}{lccccccccc}
\multicolumn{10}{c}{{\bf Spectral fits to broad-band  \emph{XMM-Newton}-\emph{INTEGRAL}/IBIS and \emph{Swift}/\emph{XRT}-\emph{INTEGRAL}/IBIS spectra}}\\
\hline    
                                              & {\bf $\Gamma$}           & {\bf N$_H^1$}                  & {\bf \emph{f$_1$}}      & {\bf N$_H^2$}                 & {\bf \emph{f$_2$}}       & {\bf E$_c$}&{\bf C}                                   & {\bf R} & {\bf $\chi^2$ (dof)} \\
                                              &                                    & {\bf (10$^{22}$cm$^{-2}$)}&                                     &{\bf (10$^{22}$cm$^{-2}$)}&                                     & {\bf (keV)}        &    &  &                               \\
\hline                      
{\bf\emph{XMM/ISGRI$^a$}}&1.67$^{+0.05}_{-0.04}$&10.96$^{+1.81}_{-1.56}$ &0.34$^{+0.03}_{-0.03}$&1.11$^{+0.12}_{-0.12}$&0.84$^{+0.02}_{-0.02}$& -    &0.73$^{+0.07}_{-0.06}$ & -&2403.6 (2629) \\
{\bf\emph{XMM/ISGRI$^b$}}&1.48$^{+0.06}_{-0.06}$&9.47$^{+2.14}_{-1.98}$&0.25$^{+0.04}_{-0.05}$&0.95$^{+0.12}_{-0.13}$&0.82$^{+0.03}_{-0.03}$&75.3$^{+25.7}_{-15.9}$& 0.85$^{+0.08}_{-0.07}$&-&2354.0 (2628)\\
{\bf\emph{XMM/ISGRI$^c$}} &1.54$^{+0.08}_{-0.08}$&8.19$^{+2.27}_{-1.92}$&0.25$^{+0.04}_{-0.05}$& 0.97$^{+0.12}_{-0.12}$&0.83$^{+0.03}_{-0.03}$&99.1$^{+54.5}_{-29.1}$& 0.67$^{+0.18}_{-0.12}$& 0.42$^{+0.47}_{-0.39}$& 2350.9 (2627)\\
{\bf\emph{XRT1/ISGRI$^d$}}&1.58$^{+0.32}_{-0.25}$& 6.62$^{+39.74}_{-6.57}$& 0.13$^{+0.08}_{-0.08}$& 1.91$^{+0.59}_{-0.49}$ & 0.75$^{+0.08}_{-0.06}$& $>$50 &0.78$^{+0.32}_{-0.32}$&-&138.8 (145)\\
{\bf\emph{XRT2/ISGRI$^d$}} & 1.75$^{+0.15}_{-0.20}$&4.95$^{+1.18}_{-1.64}$& 0.77$^{+0.06}_{-0.07}$& 0.56$^{+1.95}_{-0.13}$ & 0.95$^{+0.01}_{-0.961}$&$>$80&1.59$^{+0.56}_{-0.31}$&-& 47.8 (54)\\   
\hline
\end{tabular}
\end{center}
\scriptsize
Best fit parameters related to model model (a) (\texttt{wa*pcfabs*pcfabs*po}),  model (b) (\texttt{wa*pcfabs*pcfabs*cutoffpl}), model (c) (\texttt{wa*pcfabs*pcfabs*pexrav}) and  model (d) (\texttt{wa*pcfabs*pcfabs*cutoffpl});\\
Parameters are as following: {\bf $\Gamma$}=Photon index; {\bf N$_H^1$}= Column density of first absorber; {\bf \emph{f$_1$}}= Covering fraction of first absorber;  {\bf N$_H^2$}= Column density of second absorber;             {\bf \emph{f$_2$}}= Covering fraction of second absorber;  {\bf E$_c$}= Cut-off energy; {\bf C}=Cross calibration constant; {\bf R}= reflection Component; {\bf $\chi^2$ (dof)}= Chi squares (degrees of freedom) \\
\end{table*}

\section{Discussion and conclusions}

The radio counterpart of IGR J21247+5058 shows characteristics typical of  a Broad Line Radio Galaxy, 
probably of the FRII type. The spectrum between 610 MHz and 15 GHz,
obtained with VLA and GMRT archival data, is typical of synchrotron self-absorption radiation. Moreover, there is
evidence for a moderate Doppler boosting of the base of the jet, based on the correlation between core and total
radio power. 

The 0.4-100 keV broad-band spectrum of IGR J21247+5058 is well
modelled by a power-law continuum with a cut-off at around 70-80 keV, absorbed  by
two layers of cold material partially covering the central source.
A weak iron line is possibly present in the data, while the value of the reflection component is low (R$\sim$0.4) and not well constrained.
In many ways IGR J21247+5058 behaves like other BLRG which show weak reprocessing components and flatter X/gamma-ray
power-law slopes than generally observed in radio quiet broad line galaxies like Seyferts (\citealt{b39});
both characteristics are generally interpreted as due to the presence of a beamed jet component in these bright radio sources.
The effect of this component is that of contaminating/diluting the Seyfert continuum which is likely present in 
the source. Indeed the high energy cut-off inferred by the \emph{INTEGRAL} data is typical of Seyfert galaxies and not
of beamed AGN like blazars, which tend to have much higher cut-off energies.
We have not found strong evidence for a jet component in our broad-band data, although
we cannot exclude its presence. A better way to look for 
beamed radiation in IGR J212471+5058, is by means of high resolution observations
at radio frequencies, which would be extremely important and are therefore highly encouraged. 
On the other hand, complex absorption is not a characteristic of classical BLRG and has been observed so 
far in only another object, i.e. 3C 445 \citep{b62}. Both galaxies require cold absorbers partially covering 
the central source, even though in the case of 3C 445
three layers of obscuring materials are required while only two in IGR J21247+5058 are needed at low energies. However,  
while 3C 445 shows a strong iron line and a consequent 
strong reflection component, in the case of IGR J21247+5058 reflection is not required by the data and the iron line is 
weak or even absent.
Several emission lines are also observed at low energies in the spectrum of 3C 445, making it somewhat
different from IGR J21247+5058. As for 3C445,
the presence of absorption in IGR J21247+5058 is at odds with its Seyfert 1 classification, but this discrepancy 
can be circumvented if the {\textquotedblleft{clumpy torus}\textquotedblright} model 
recently proposed by \citet{b66} and successfully applied to 3C 445 \citep{b62} is also applied to this newly discovered radio galaxy. In this model, 
the torus is not a continuous toroidal structure but is made of clouds with N$_H$ $\sim$ 10$^{22}$-10$^{23}$
atoms cm$^{-2}$ distributed around the equatorial plane of the AGN. The BLR represents the inner segment of the torus
and this is the region where the X-ray absorption is likely to occur \citep{b67}. Because of this clumpiness
the difference between type 1 and 2 AGN is not only due to orientation but also to the probability 
of direct view of the active nucleus or, in other words,
on how many clouds our line of sight intercepts.  Indeed, the inclination to the line of sight inferred from the radio analysis 
suggests that we are peering directly into the broad line
region; moreover, variations in the absorption properties are expected in this scenario, as the torus structure changes due to cloud motion.

Overall we can conclude that IGR J21247+5058 is a new interesting member of the class of BLRG, with features that are 
in some respect typical of this class and others which are quite rare for its type: its brightness and proximity 
make it an ideal laboratory in which to study radio galaxies properties.



\label{lastpage}
\end{document}